\documentclass[smallabstract,smallcaptions]{dccpaper}

\usepackage[utf8]{inputenc}

\usepackage[T1]{fontenc}

\usepackage{amsmath}

\usepackage{amssymb}

\usepackage{url}

\usepackage{graphicx}

\usepackage{booktabs}

\usepackage{tabularx}

\begin{document}

\title{\large\textbf{A Configurable ETC2 Encoder with Learned Search Allocation}}

\author{Ziyu Guo, Haomiao Jiang, Sergey Makeev, Angelo Pesce \\
Roblox \\
\texttt{\{zguo, haomiaojiang, smakeev, apesce\}@roblox.com}}

\Notes{This work has been submitted to the IEEE for possible publication. Copyright may be transferred without notice, after which this version may no longer be accessible.}

\maketitle

\thispagestyle{empty}

\begin{abstract}

ETC2 fixes the texture representation and bit rate, leaving the encoder to trade search time for distortion. We present a configurable CPU encoder in which rules or small neural networks allocate conventional ETC2 searches; these searches construct every encoded block. Configurations span fast-encoder output through broader mode and endpoint searches. On 64 color textures used to evaluate and select configurations, the fastest measured configurations reaching at least Google Etc2Comp's pooled REC709 PSNR at efforts 0--49.5 run 30.7--40.9\(\times\) faster. A high-quality configuration exceeds effort-100 PSNR with 64.1\% less time. At equal search quotas, learned action ranking and second-search allocation jointly gain 0.124 dB over rule-based choices. At high quality, learning saves time at a small distortion cost. Outputs retain standard ETC2 storage and hardware decoding.

\end{abstract}

\Section{1. Introduction and related work}

ETC2 stores each 4\(\times\)4 color block in eight bytes, supporting hardware decoding and random access \cite{r1}. With the rate fixed, encoding trades distortion against computation. Interactive previews and offline conversion need different tradeoffs: on our 64-texture set, etcpak takes under a second and Google's Etc2Comp \cite{r11} takes hundreds to tens of thousands of seconds.

Additional search is not equally useful for every block. Its benefit depends on the candidate representation, and its cost depends on how much work can be pruned after earlier results. A block with high current error need not benefit most from another search. We study how to allocate this computation: rules and small neural networks choose which blocks to search, which searches to run and how thoroughly to run them. Conventional encoder code constructs the compressed blocks.

This follows learned search restriction in video encoding. Galpin et al. \cite{r6} combine learned partition preselection with encoder search restrictions and adjustable thresholds; Tech et al. \cite{r7} learn VVC restrictions from rate--distortion--time costs and include inference in their comparisons. ETC2 has no rate decision, so our corresponding objective is distortion versus encoding time. TexNN \cite{r5} uses learned configuration prediction to accelerate ASTC encoding; learned BC7 mode selection \cite{r8} similarly restricts conventional search. In contrast, neural block fitting \cite{r9} generates block-format data from per-material networks, and neural material representations \cite{r10} replace conventional formats with learned decoders. Our networks only allocate encoder computation.

QuickETC2 \cite{r2} and QuickETC2-HQ \cite{r3} accelerate ETC2 encoding through luma approximations, mode trials and more accurate error comparisons. We incorporate etcpak \cite{r13}, which includes QuickETC2 techniques, and Nah's QuickETC2-HQ implementation as fast initial encoders. H-ETC2 \cite{r4} combines an etcpak CPU encoding with error-selected GPU refinement. Here both search and inference run on the CPU, with configurations covering fast conversion through quality slightly above Etc2Comp effort 100.

Our contribution is this configurable implementation and an analysis of which decisions benefit from learning. Low budgets favor avoiding full-image analysis and data movement. At intermediate budgets, learned search choices improve quality, while a simple rule for limiting modifier-table trials is competitive with learning. At high quality, learned choices save time with a small quality loss, but expensive searches remain difficult to skip. Sections 4 and 5 test these findings through whole-encoder comparisons and controlled changes to search decisions.

\Section{2. Search allocation and implementation}

Individual and Differential modes split a block into left/right or top/bottom halves, its \emph{orientation}. Each half has a base color, or \emph{endpoint}, and one of eight modifier tables ordered by increasing amplitude. Each pixel's two-bit selector chooses one of the table's four luma offsets. Individual stores two four-bit-per-channel endpoints; Differential stores a five-bit-per-channel base with three-bit deltas, restricting their separation. T and H construct four-color palettes; Planar represents a smoothly varying surface.

We start from a valid initial encoding, the \emph{seed}. An \emph{action} is a conventional search for a specified mode, orientation and endpoint neighborhood. The twelve-action menu contains Differential/Individual \(\times\) two orientations \(\times\) radii zero/one (eight), two Planar radii, and separate T and H actions at radius zero. For Differential and Individual, radius one searches within one quantization step per channel of the half-block means. The \emph{table budget} limits trials to the first two, four, six or eight modifier tables. Native C++ searches adapted from Etc2Comp construct and score the candidates. A configuration specifies the seed, action menu, block quotas, table budgets, selection rules or trained models, and acceptance test.

\SubSection{2.1 From an initial encoding to a second search}

The six-candidate seed evaluates Differential and Individual in both orientations, Planar, and T-or-H through an expanded QuickETC2 path, selecting the lowest decoded REC709 error (Section 3). This unrepaired seed takes about 7.2 s on the evaluation set. Its constant-block shortcut can miss the source color; repaired configurations search single-color encodings before allocating further work.

Admission selects blocks by seed error; a source ranker orders their actions. The first search returns an encoding, errors and the attained action: an invalid H request can fall back to Differential. The next proposal is the highest-ranked action distinct from the requested and attained actions. \emph{Continuation} selects blocks for this second search, starting from the first encoding (Figure 1).

\begin{figure}[tbp]
\centering
\includegraphics[width=\linewidth]{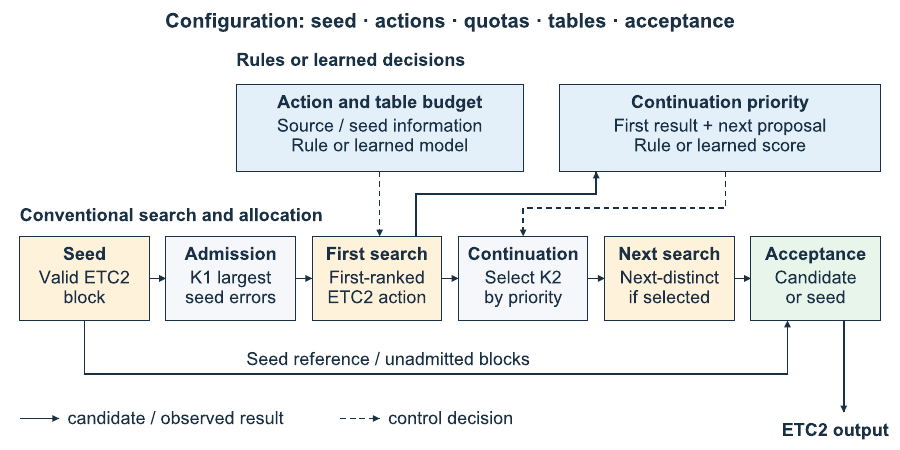}
\caption{Two-action path. Solid arrows carry encoded candidates or observed results; dashed arrows apply search decisions. First-search quota $K_1$ is per image or pooled over four images; continuation quota $K_2$ is per image. Acceptance compares the final candidate with the seed.}
\label{fig:generalized-1}
\end{figure}

Fractions $f_1,f_2$ are measured against all blocks, not only those previously selected. For per-image allocation, image $i$ has $N_i$ blocks and $N_{e,i}$ eligible positive-error blocks:

\[
K_{1,i}=\min\{\lfloor f_1N_i\rfloor,N_{e,i}\},\qquad
K_{2,i}=\min\{\lfloor f_2N_i\rfloor,K_{1,i}\}.
\]

Thus per-image 60/30 allocation first searches up to 60\% of each image and continues up to 30\%. \emph{Job-pooled admission} allocates the first quota across a fixed four-image batch by seed error, allowing different coverage per image. Continuation remains image-local and unused quotas are not transferred. The headline configurations use job-pooled admission; the learning comparisons use per-image admission.

Strict one-action configurations accept lower rounded, live-decoded REC709 cost. Since REC709 improvements can increase RGB error, color-gated configurations compare the final candidate $c$ with the seed $0$, requiring $S_c<S_0$ and, for positive errors, $\log(S_c/S_0)+0.265\log(R_c/R_0)\le0$. Here $S$ is the encoder's internal REC709 selection cost and $R$ is RGB squared error; the gate permits an RGB increase only when compensated by REC709 improvement. The coefficient was fixed on earlier development data. Internal selection uses unrounded seed colors and live candidate costs, with continued T/H candidates rescored from packed blocks. Rejection returns the seed. These internal costs are distinct from the independently decoded errors reported in Section 3.

\SubSection{2.2 Choosing actions and table budgets}

The source ranker uses source pixels and descriptors of variation, orientations, gradients, Planar residuals and endpoint validity. For a Differential/Individual seed, \emph{seed-based ranking} first tries its mode and orientation at radius one, then the other mode in that orientation, then the original mode in the other orientation. T/H seeds try their own radius-zero action, then the other; Planar starts with radius-one Planar. Remaining radius-one actions precede T, H and radius-zero actions, without duplicates.

The learned continuation head scores the next proposal from the ranker's hidden features and scores, the attained and proposed actions, and observed first-search errors. The highest priorities fill the continuation quota. \emph{Attained-error continuation} instead gives the second search to blocks with the largest remaining REC709 error.

Table budgets control work within an action. Adaptive budgets apply to Differential radius-one searches; Individual retains eight tables. The learned table-budget head predicts a two/four/six/eight-table prefix for each half. A nonlearned \emph{source-spread rule} takes the maximum absolute deviation of eight-bit REC709 luma from its half-block mean, over both oriented halves. Deviations at most 17, 42 or 80 select two, four or six tables for both halves; larger deviations select eight. These thresholds are the largest modifier amplitudes in the corresponding prefixes. Search kernels also prune endpoint candidates whose lower bounds cannot improve the current result, so table counts alone do not determine runtime.

\SubSection{2.3 Fast and high-quality configurations}

At the fast end, configurations return etcpak, its variant without heuristic mode selection, or QuickETC2-HQ output directly. These encoders expect BGRA; converting RGBA input during block gathering avoids a full-image temporary without changing the bytes. No decoding, sorting or inference is needed.

Cross-encoder refinement tries the other fast encoder's candidate; fixed-palette refinement only reselects pixel selectors. Each stage accepts lower decoded REC709 error with no RGB-error increase. A deterministic hash of block index selects coverage, with larger thresholds admitting supersets. Eligibility can be limited to Individual-mode seeds, where most cross-encoder gain occurs, plus constant blocks that this mode test misses. \emph{Sparse native refinement} starts from cross-refined etcpak output and packs only hash-admitted blocks for scoring and one rule- or model-selected action at a fixed table budget. Feature extraction and inference therefore scale with admitted coverage.

At high quality, \emph{menu search} starts from the repaired six-candidate seed and runs the leading three, four or eight ranked actions on every block at eight-table budgets. Eight-action configurations can then apply \emph{offset refinement}: four groups of Differential radius-one searches centered on half-block means shifted by gray offsets, optionally preceded by radius-one T/H searches. \emph{Residual-error admission} ranks blocks within each image by current REC709 error; \emph{learned-gain admission} ranks by predicted improvement. Each selects a specified fraction for six- or eight-table offset refinement. Each operation retains the best decoded REC709 candidate. Exact-bound kernels skip provably unhelpful work with unchanged output. The \emph{eight-table reference} runs eight learned-ranked menu actions, then T/H and all four offset groups at eight tables on every block.

\SubSection{2.4 Models and supervision}

Four learned functions implement these decisions: source ranking, continuation priority, table budget and offset-refinement gain. Source ranking learns to prefer cheap, near-best actions. Offline encoding supplies outcomes for all twelve actions. Among valid actions within 1\% of the REC709 winner's REC709 and RGB errors, the target minimizes endpoint-pair trials, a search-work proxy. Regression, ranking and relative-error losses favor that target. Continuation fits gain and validity targets for the second search and distills priorities from a pretrained controller. Table-budget targets are the shortest prefixes containing the tables selected by eight-table searches for each half. The offset-gain predictor learns retained REC709 improvement from source features and the current action and errors, fitting on eight development families and selecting its checkpoint on four others.

The compact source model receives 48 normalized RGB samples and 64 descriptors. Its \texttt{112\(\to\)64\(\to\)64} ReLU trunk produces twelve scores and 64 hidden features; source and continuation together have 23,579 FP32 parameters. The optional table-budget head adds 5,848. We also evaluate a 160,235-parameter source/continuation model with width 192 and two residual units; the controlled learning comparisons use the compact model. Sparse native refinement invokes only the compact source ranker; two-action paths also invoke continuation. The separate offset-gain network is \texttt{126\(\to\)64\(\to\)32\(\to\)1}.

The compact model trains on one million color blocks from 5,087 images. AdamW uses learning rate 0.0002, batches of 4,096 and 20 weighted-resampling epochs. Epoch 11 minimizes a quality and search-work penalty on separate model-selection data. Table-budget fitting and checkpoint selection also use disjoint groups.

\Section{3. Evaluation design}

We evaluate opaque color/albedo textures from AmbientCG and Poly Haven. Our evaluation set contains one image from each of 64 material families (a material and its lettered or numbered variants), totaling 66.06 megapixels. The first 32 eligible families per source in seeded hash order were selected without codec outcomes. Recorded training and checkpoint-selection families are excluded, with duplicate checks before encoding. We also used this set to select encoder configurations.

Quality is pooled PSNR in Etc2Comp's REC709 metric \cite{r11}, which Etc2Comp also optimizes here. For RGB8 channel residuals divided by 255, define $Y=.2126d_R+.7152d_G+.0722d_B$, $C_r=(d_R-Y)/(2(1-.2126))$ and $C_b=(d_B-Y)/(2(1-.0722))$. The error is $E_{709}=\sum(3Y^2+C_r^2+C_b^2/2)$, and PSNR is $10\log_{10}(P/E_{709})$ for $P$ pixels. Errors are summed across images before conversion to decibels, without transfer-function linearization. Reporting uses independently decoded ETC2 bytes, not the encoder's internal selection costs.

Single-thread measurements use a pinned Xeon Platinum 8259CL core. The warmed in-memory timer covers RGBA8 input through materialized ETC2 output, including conversion, allocation, CPU inference and refinement; loading, file I/O and independent scoring are excluded. The sixteen four-image allocation jobs remain fixed. We compare etcpak, QuickETC2-HQ, fast-search ETCPACK 2.74 \cite{r14} in RGB/perceptual modes, and public Google Etc2Comp at efforts 0, 10, 25, 49.5, 75 and 100. Etc2Comp refines high-error blocks toward an effort-dependent completion fraction; above 49.5 it adds radius-one T/H, then progressively shifted Differential groups. Effort 49.5 stops after radius-one Planar, before those additions. The build uses GCC 12, \texttt{-O3} and AVX2 with fast-math and contraction disabled. Our Etc2Comp-derived menu search also disables contraction; fast-encoder and optimized offset kernels use FMA.

Table 1's rule-based point, the intermediate comparisons, ETCPACK and the direct and refined fast paths use arithmetic means of three randomized rounds. Table 1's learned points, Google Etc2Comp and the high-quality configurations use single timings. Table 2's four configurations share one randomized session, also supplying their Figure 2 points. Repeated rounds are averaged within each measurement session. Byte comparisons verify integration. We plan to release the implementation and evaluation tools as open source.

The experiments ask where search allocation matters. Whole-encoder comparisons vary seeds, coverage, action counts and table budgets. Intermediate comparisons hold the repaired six-candidate seed, 60/30 per-image quotas, search kernels and acceptance fixed, replacing learned ranking/continuation jointly by seed-based ranking/attained-error continuation, then varying table budgets. High-quality comparisons cross action ranking with offset-refinement admission at fixed coverage. Recorded-outcome \emph{replays} recombine completed searches; \emph{oracles} choose using those known outcomes to estimate potential gains. Sampled diagnostics identify their separate development populations below.

\Section{4. End-to-end evaluations}

\begin{figure}[!t]
\centering
\includegraphics[width=\linewidth]{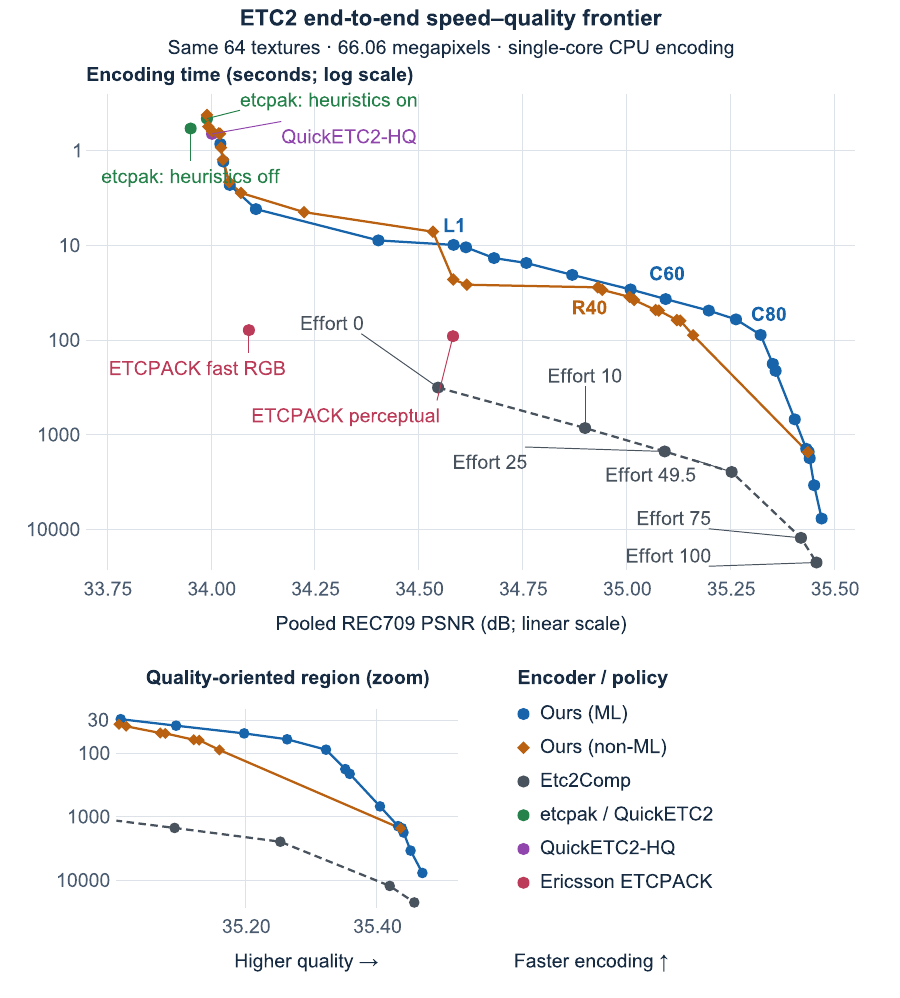}
\caption{Results on the 64-texture evaluation set; each point is a measured configuration. Higher PSNR is rightward and faster encoding upward. Blue circles use learning; orange diamonds do not. Labels L1, R40, C60 and C80 identify Table 1. Curves are filtered separately within these two categories, omitting dominated points and slower near-ties within 0.002 dB. This tolerance controls display density. All external settings are shown.}
\label{fig:generalized-2}
\end{figure}

\textbf{Fast encoding.} Fused conversion cuts etcpak from 0.458 to 0.418 s and QuickETC2-HQ from 0.664 to 0.623 s, saving 8.62\% and 6.20\% with identical output. Cross-encoder refinement of etcpak's Individual-mode and constant blocks reaches 34.022 dB in 0.662 s, 0.019 dB above the original QuickETC2-HQ at similar time. From this seed, sparse native refinement with learned ranking spans 0.85--50.85 s as hash-admitted coverage rises from 0.1\% to 100\%.

\textbf{Quality-matched comparisons.} Among the 138 measured internal configurations, Table 1 reports the fastest whose pooled REC709 PSNR meets each named Etc2Comp target. All four use the six-candidate seed and job-pooled first-search allocation. Rule-only R40 includes constant-block repair and the color-error gate; L1 and C60 use strict one-action acceptance. C80 adds learned continuation and the color-error gate.

\begin{table}[tbp]
\centering
\caption{Fastest measured configurations reaching each Etc2Comp reference PSNR. L/C denote larger/compact learned rankers; R denotes seed-based ranking. The number is first-search coverage in percent; only C80 continues, on 40\% of each image. Learned configurations use learned Differential budgets and eight Individual tables; R40 uses four/eight Differential/Individual tables. PSNR is in dB and time in seconds.}
\label{tab:generalized-1}
{\small
\begin{tabularx}{\linewidth}{@{}Xrrrrrr@{}}
\toprule
ID & Effort & Ref. dB & Ref. s & Ours dB & Ours s & Speedup \\
\midrule
L1 & 0 & 34.5467 & 317.96 & 34.5839 & 9.88 & 32.18\(\times\) \\
R40 & 10 & 34.9004 & 851.50 & 34.9311 & 27.75 & 30.68\(\times\) \\
C60 & 25 & 35.0921 & 1504.18 & 35.0944 & 36.93 & 40.73\(\times\) \\
C80 & 49.5 & 35.2532 & 2478.65 & 35.2636 & 60.53 & 40.95\(\times\) \\
\bottomrule
\end{tabularx}}
\end{table}

Etc2Comp spans the widest measured external quality range. ETCPACK perceptual is faster and higher-quality than Etc2Comp effort 0; L1 reaches at least its PSNR at 9.22\(\times\) its speed, versus 32.18\(\times\) relative to Etc2Comp effort 0.

\textbf{High quality.} Four-action eight-table menu search reaches 35.359 dB in 211.76 s. The eight-action configuration with learned ranking and 25\% learned-gain offset admission takes 1409.56 s at 35.4327 dB, the fastest measured configuration reaching effort-75 PSNR (35.4200 dB, 12303.89 s), at 8.73\(\times\) its speed. The eight-table reference takes 8045.02 s versus 22419.59 s for Etc2Comp effort 100, saving 64.12\% while increasing PSNR from 35.4574 to 35.4696 dB. The scheduler in Section 5 reaches 35.4697 dB in 7693.86 s; Figure 2 displays this measured point.

\Section{5. Analysis of search allocation}

\textbf{Sparse processing and admission.} At 1\% hash admission, one compact-ranked, four-table action from the cross-refined etcpak seed reaches 34.030 dB in 1.30 s, including 0.08 s for ranking. Processing only admitted blocks is inexpensive, but this point remains below effort-0 quality; L1 reaches that target using the six-candidate seed.

To test admission signals, we replay full-coverage results on the evaluation set, keeping refinement for 1\% of each image's blocks and the seed elsewhere. Gains over the seed are 0.0079 dB for hash admission, 0.0213 dB for largest source channel range, 0.0514 dB for largest decoded seed error and 0.0836 dB for an oracle using known gain. These equal-count replays exclude signal-computation costs; the deployed hash needs neither decoding nor sorting. At full one-action coverage, learned ranking gains 0.0655 dB over seed-based ranking for 8.03\% more encoding time, including inference.

\textbf{Learning at intermediate budgets.} With shared source-spread table budgets and 60/30 per-image quotas, learned ranking and continuation take 50.58 s versus 51.67 s for seed-based ranking and attained-error continuation, jointly gaining 0.1239 dB. Replacing the source-spread rule with learned table budgets adds only 0.0008 dB, at 50.29 s.

On 24 development images, shuffling learned budgets among blocks sharing image, action and orientation preserves budget distributions but loses 0.035--0.039 dB with less than 1\% timing change. Content-dependent assignment matters, but codec structure supplies a simple rule. In a replay of 1,459 admitted Differential radius-one blocks from those images, oracle prefixes preserve first-search output while removing 31.2\% of table trials versus learned budgets, but only 6.5\% of lower-bound computations. Fewer table trials need not yield proportional runtime savings.

\textbf{High-quality decisions.} Table 2 fixes eight menu actions, eight-table menu budgets and 25\% offset-refinement coverage. Ranking changes both which actions run and their order; admission changes which blocks receive offset refinement. Learned ranking lowers measured time by 6.40--7.22\% with about 0.0058 dB lower PSNR; learned admission adds about 0.0019 dB under either ranking. Together they save 7.86\% with a 0.00385 dB reduction. The rule-only configuration also reaches high quality.

\begin{table}[tbp]
\centering
\caption{High-quality factorial on the 64-texture evaluation set: eight menu actions at full coverage and eight-table budgets, then six-table offset refinement of 25\% of each image. One randomized timing session supplies this table and the corresponding Figure 2 points.}
\label{tab:generalized-2}
{\small
\begin{tabularx}{\linewidth}{@{}XXrX@{}}
\toprule
Action ranking & Offset admission & Seconds & REC709 PSNR dB \\
\midrule
Seed-based & Residual error & 1529.87 & 35.4365 \\
Seed-based & Learned gain & 1519.30 & 35.4385 \\
Learned & Residual error & 1432.00 & 35.4307 \\
Learned & Learned gain & 1409.56 & 35.4327 \\
\bottomrule
\end{tabularx}}
\end{table}

Learned ranking reduces menu-search time from about 531 s to 415--418 s with roughly 33 million executed actions in either case. Equal action counts do not imply equal work: action identity, order and incumbent-dependent pruning change cost. The ranking phase rises from under one second for rules to about eleven seconds with inference, while offset refinement takes 946--966 s. This shared stage limits the total saving. Exact-bound kernels reduce the eight-table reference from 10330.00 to 8045.02 s, saving 22.12\% with unchanged output.

\textbf{Can offset refinement be skipped?} On 24,576 uniformly sampled blocks from twelve development images outside the evaluation set, we replay fifteen bundles: no refinement; T, H or T/H; each offset group alone; T/H plus groups 1, 1--2 or 1--3; and all four groups at six or eight tables, with or without T/H. Other bundles use eight tables. A bundle is \emph{adequate} if its result is no worse in REC709 and RGB error than T/H plus all four groups at eight tables. This reference refinement spends 87.34\% of search time on blocks with no retained improvement. An oracle choosing each block's cheapest adequate bundle uses 2.073\% of reference refinement time, excluding base encoding and bundle selection.

We predict each bundle's gain, cost and inadequacy using 16,384 fitting blocks and 8,192 validation blocks from disjoint development families. Inputs extend the offset-admission features with the current encoded block. The scheduler chooses the lowest predicted-cost bundle with inadequacy score at most 0.001, always allowing the reference bundle. It retains T/H and all four offset groups on 99.9\% of evaluation-set blocks, using six tables on 46.6\% and eight on 53.3\%, then tries the four remaining menu actions on blocks with nonzero error. On validation families, oracle costs barely change its choices, whereas oracle adequacy permits much cheaper search. Predicting when a cheaper bundle suffices is the limiting decision.

A second scheduler predicts quality loss relative to the reference refinement. It reaches 35.4392 dB in 1949.65 s; fixed 40\% learned-gain offset admission reaches 35.4410 dB in 1775.50 s. Direct coverage control gives the better measured tradeoff.

\textbf{Perceptual-metric check.} The fully learned 60/30 configuration has lower FLIP error \cite{r12} on 52 of 64 images than the repaired, seed-ranked, eight-table 100/100 configuration, taking 50.29 versus 115.33 s. This compares operating points with different quotas. These findings concern color textures; normal maps, roughness and packed channels require different distortion objectives.

\Section{6. Conclusion}

Allocating ETC2 searches covers a wide range of encoding times without changing the representation or decoder. On the evaluation set, the fastest measured configurations reaching Etc2Comp efforts 0--49.5 in pooled REC709 PSNR are 30.7--40.9\(\times\) faster; the eight-table reference exceeds effort-100 PSNR with 64.1\% less time.

Joint action ranking and continuation improve intermediate-budget quality by about 0.124 dB over rule-based choices, while a codec-derived table-budget rule nearly matches learned budgets. High quality is also attainable without learning; learned ranking mainly reduces time at a small quality cost. A next step is to predict which offset-refinement searches can be skipped while preserving reference quality.

\textbf{AI assistance.} OpenAI Codex assisted code implementation, experiment automation, data analysis and manuscript preparation.

\Section{References}


\begin{thebibliography}{14}

\bibitem{r1} The Khronos Group. \emph{OpenGL ES Version 3.0.6}, \S\S~3.8.6, C.1, C.1.1. 2019.

\bibitem{r2} J.-H. Nah. QuickETC2: Fast ETC2 Texture Compression using Luma Differences. \emph{ACM Transactions on Graphics} 39(6), article 270, 2020. DOI: 10.1145/3414685.3417787.

\bibitem{r3} J.-H. Nah. QuickETC2-HQ: Improved ETC2 Encoding Techniques for Real-Time, High-Quality Texture Compression. \emph{Computers \& Graphics} 116, 308--316, 2023. DOI: 10.1016/j.cag.2023.08.032.

\bibitem{r4} H.-k. Lee and J.-H. Nah. H-ETC2: Design of a CPU-GPU Hybrid ETC2 Encoder. \emph{Computer Graphics Forum} 42(7), e14969, 2023. DOI: 10.1111/cgf.14969.

\bibitem{r5} S. Pratapa, T. Olson, A. Chalfin, and D. Manocha. TexNN: Fast Texture Encoding Using Neural Networks. \emph{Computer Graphics Forum} 38(1), 328--339, 2019. DOI: 10.1111/cgf.13534.

\bibitem{r6} F. Galpin, F. Racapé, S. Jaiswal, P. Bordes, F. Le Léannec, and E. François. CNN-based driving of block partitioning for intra slices encoding. \emph{Data Compression Conference}, 2019. DOI: 10.1109/DCC.2019.00024. Author version: arXiv:2011.06691.

\bibitem{r7} G. Tech, J. Pfaff, H. Schwarz, P. Helle, A. Wieckowski, D. Marpe, and T. Wiegand. Fast Partitioning for VVC Intra-Picture Encoding with a CNN Minimizing the Rate-Distortion-Time Cost. \emph{Data Compression Conference}, 3--12, 2021. DOI: 10.1109/DCC50243.2021.00008.

\bibitem{r8} T. Madams. Optimized Compression Mode Selection for BC7 Texture Encoding. US Patent Application Publication 2025/0014231 A1, 2025.

\bibitem{r9} S. Fujieda and T. Harada. \emph{Neural Texture Block Compression}. arXiv:2407.09543, 2024.

\bibitem{r10} K. Vaidyanathan, M. Salvi, B. Wronski, T. Akenine-Möller, P. Ebelin, and A. Lefohn. Random-Access Neural Compression of Material Textures. \emph{ACM Transactions on Graphics} 42(4), 2023. DOI: 10.1145/3592407.

\bibitem{r11} Etc2Comp Authors. \emph{Etc2Comp: Texture to ETC2 Compressor}. Software, commit 39422c1. \url{https://github.com/google/etc2comp}.

\bibitem{r12} P. Andersson, J. Nilsson, T. Akenine-Möller, M. Oskarsson, K. Åström, and M. D. Fairchild. FLIP: A Difference Evaluator for Alternating Images. \emph{Proceedings of the ACM on Computer Graphics and Interactive Techniques} 3(2), 1--23, 2020. DOI: 10.1145/3406183.

\bibitem{r13} B. Taudul. \emph{etcpak}. Software, commit 3d716e1. \url{https://github.com/wolfpld/etcpak}.

\bibitem{r14} Ericsson AB. \emph{ETCPACK}, version 2.74. Software.

\end{thebibliography}
\end{document}